# Critical Current Survival in YBCO Superconducting Layer of the Delaminated Coated Conductor

Feng Feng[1], Qishu Fu[1], Timing Qu[2]*, Chen Gu[3], Yubin Yue[4], Hui Mu[1], Xiangsong Zhang[1], Hongyuan Lu[1], Linli Wang[1], Siwei Chen[3,4] and Pingfa Feng[1]

1. Division of Advanced Manufacturing, Graduate School at Shenzhen, Tsinghua University, Shenzhen 518055, China
2. Department of Mechanical Engineering, Tsinghua University, Beijing 100084, China
3. Department of Physics, Tsinghua University, Beijing 100084, China
4. Beijing Eastforce Superconducting Technology Co., Ltd., Beijing 100085, China.

E-mail: tmqu@mail.tsinghua.edu.cn

**Abstract**

High temperature superconducting coated conductor (CC) could be practically applied in electric equipment due to its favorable mechanical properties and the critical current performance of YBCO superconducting layer. It is well known that CC could be easily delaminated because of its poor stress tolerance in thickness direction, i.e. along the c-axis of YBCO. Commonly, a stack including YBCO layer and silver stabilizer could be obtained after the delamination. It would be interesting to investigate the superconducting properties of the delaminated stack, since it could also be considered as a new type of CC with the silver stabilizer as the buffer layer, which is quite different from the oxide buffer layers in the traditional CC and might lead to new applications. In this study, a CC sample was delaminated by liquid nitrogen immersing. A Hall probe scanning system was employed to measure the critical current ($I_C$) distribution of the original sample and the obtained stack. It was found that $I_C$ could be partially preserved after the delamination. Dense and crack-free morphologies of the delaminated surfaces were observed by scanning electron microscopy, and the potential application of the obtained stack in superconducting joint technology was discussed.

High temperature superconducting (HTS) coated conductor (CC), as known as second generation HTS wire, is generally regarded as the main superconducting candidate for various electric devices [1, 2, 3]. The nickel-based substrate could provide favorable mechanical strength in the length direction and width direction of CC. However, in the thickness direction, i.e. along the *c*-axis of YBCO, the adhesion strength of layer interfaces is weak [4, 5], which could lead to the delamination behavior. The stress tolerance in the thickness direction, might be exceeded in many cases, such as thermal cycling [6, 7], Lorentz force [8, 9], epoxy impregnation [10], hoop stresses [11, 12], etc. The delamination is a severe threat for CC practical applications [13, 14].

Therefore, CC delamination was investigated by many groups. Takematsu et al. studied the structural changes of an epoxy impregnated double pancake coils after five times temperature cycling of 77 K to room temperature [7]. Van et al. used two anvils to apply tensile strength on both sides of CC and detect the delamination strength, which was about 10 MPa for the 8μm-thick YBCO layer [4]. Kesgin et al. used the T peel test to delaminate CC samples, and the delamination force and surface topography were measured. It was concluded that the delamination could occur between the HTS layer and buffer layer or inside the HTS layer [15].

On the other hand, delamination could be utilized to fabricate ceramic thin films on plastic substrates. In the study of Kozuka et al., a stack of plastic/ceramic film/release-layer/ Si (100)-substrate was fabricated, and then delaminated to obtain the plastic/ceramic-film part [16, 17]. Such a method was proposed because the plastic substrate could not withstand the 500 °C heat treatment of ceramic film fabrication, and the delamination process did not damage the ceramic thin films, although it was a brittle material. YBCO is mechanically of ceramic nature [18], thus it could be speculated that the superconducting property might be totally or partially preserved after delamination. If the delaminated part with YBCO/silver component could carry superconducting current, it could be considered as a new type of CC with the silver stabilizer as the new buffer layer. Silver is quite different from the oxide cap layers in the traditional CC, thus the delaminated part might be used in some new applications.

Therefore, it would be interesting to study if critical current ($I_C$) of YBCO layer could preserved after delamination. In this study, a CC sample was delaminated, $I_C$ distributions of the original sample and the delaminated sample were measured using a Hall probe scanning system, and the morphologies of the delaminated surfaces were observed. $I_C$ was found to be partially preserved after delamination, and its potential application in the superconducting joint would be discussed through the estimation of oxygen diffusion durations.

The HTS CC sample used in this study was produced by AMSC Company, which owned a Ni-W RABiTS substrate about 75 μm thick, the ($Y_2O_3$/YSZ/$CeO_2$) buffer layers about 225 nm thick, a YBCO layer about 1 μm thick and a silver stabilizer about 3 μm thick. Both sides of CC were covered by brass tapes. The thickness of CC was 0.40 mm, the width was 4.40 mm, and the labeled $I_C$ was 90 A (77 K, self-field). The original sample about 60 mm long was cut out of the long CC tape.

In order to delaminate the original sample, two factors found in our preliminary experiments were considered. First, the direct delamination by peeling force could cause micro-cracks in the YBCO layer, which might be caused by the local curve which exceeded the radius limitation of CC [19]. Second, the YBCO layer near the edges was difficult to delaminate, because the adhesion strength was larger there [15]. Therefore, the delamination operation was conducted at room temperature as following: Both edges of the original sample were sheared, with the CC width decreasing to 3.63 mm; The sheared sample were placed in liquid nitrogen, and could delaminate by itself, similar to the report of Takematsu et al. [7]. The surfaces of two delaminated parts were black and purple, respectively, indicating that the delamination occurred generally at the YBCO-buffer interface. The delaminated part with YBCO layer would be denoted as "delaminated sample" in this paper.

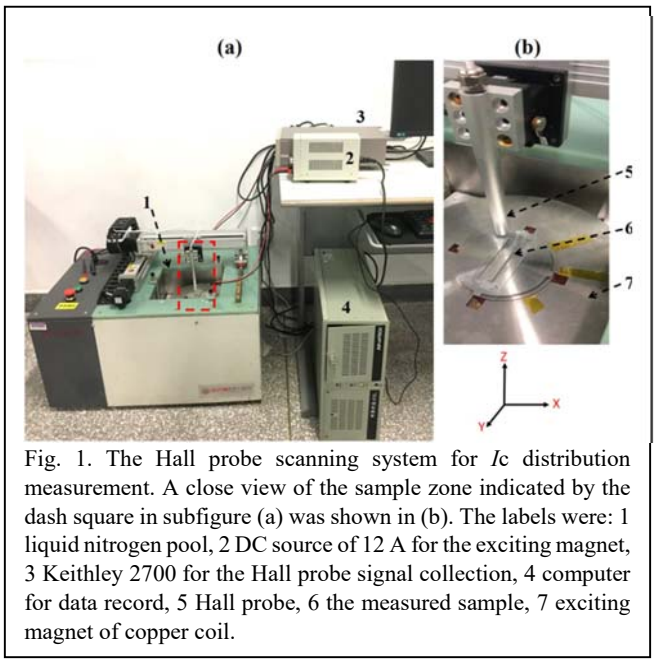

Fig. 1. The Hall probe scanning system for $I_C$ distribution measurement. A close view of the sample zone indicated by the dash square in subfigure (a) was shown in (b). The labels were: 1 liquid nitrogen pool, 2 DC source of 12 A for the exciting magnet, 3 Keithley 2700 for the Hall probe signal collection, 4 computer for data record, 5 Hall probe, 6 the measured sample, 7 exciting magnet of copper coil.

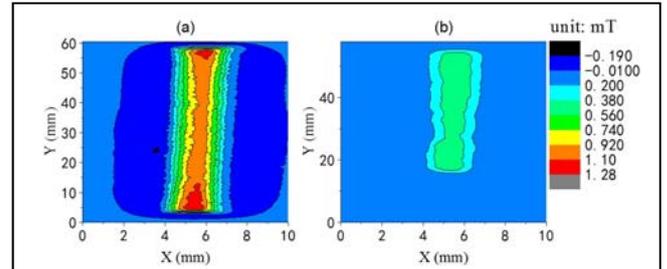

Fig. 2. The magnetic field distributions of (a) the original sample and (b) the delaminated sample, which were measured using the Hall probe scanning system.

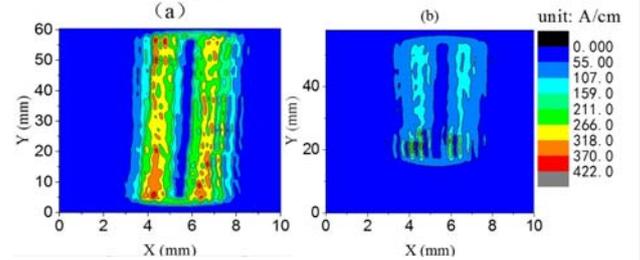

Fig. 3. The $I_C$ distributions of (a) the original sample and (b) the delaminated sample, which were calculated according to the data shown in Figure 2.

The $I_C$ distribution was measured using a proto-type Hall probe scanning system [20, 21] established by Beijing Eastforce company, as shown in Figure 1. The sample was firstly placed in the liquid nitrogen pool and cooled down to 77 K, then excited by an external magnetic field larger than 2-fold penetration field [22-24]. After switching off the exciting magnet, there was a current loop with the critical current density. The magnetic field generated by the current loop was then measured by the Hall probe 2-dimensional scanning. The $I_C$ distribution inside the YBCO layer was then calculated according to the magnetic field distribution.

The morphologies of the delaminated surfaces were observed by scanning electron microscopy (SEM) using Hitachi SU8010 with an energy dispersive spectrometer (DES) to analyze the elemental composition.

The magnetic field distributions of the original sample and delaminated sample were illustrated in Figure 2 (a) and (b), respectively. Their calculated $I_C$ distributions were shown in Figure 3 (a) and (b). $I_C$ was low in the central zones of both samples because of the field penetration mechanism, which was consisted with the report by Higashikawa et al. [20]. As could be observed in Figure 3, about 1/3 of the delaminated sample was of zero $I_C$. In the other 2/3 part, the average $I_C$ percentage relative to the original sample was about 45%. It could be certain that $I_C$ could be partially preserved in YBCO layer after delamination.

The SEM images of the delaminated sample were shown in Figure 4. The porous morphology was common for the YBCO thin films fabricated via the chemical solution deposition (CSD) method [25]. The delaminated surface was flat and dense for both zones with or without preserved $I_C$. Thus detailed reason of the zero $I_C$ zone required further investigation other than SEM. On the surface of the delaminated part with buffer layers, there was also some porous morphology, indicating the existence of YBCO grains. The EDS result shown in Figure 4 (d) could verify that some YBCO grains were attached on the buffer layer. Therefore, the delamination occurred inside the YBCO layer.

As mentioned in the above section, it could be concluded that the delaminated sample was a YBCO/Ag/Brass stack, as shown in Figure 5 (a). Such a stack could be considered as a new type of coated conductor with the silver stabilizer as its buffer layer. Silver owned a large oxygen diffusion coefficient ($D$) of $10^{-6}$ cm$^2$/s at 450°C~600°C (a typical oxygenation temperature range) [26], which is much larger than YBCO. As reported by Park et al. [27], there was a main difficulty of superconducting joint for CC, that the oxygenation process would cost a very long duration due to the low $D$ value in YBCO and long diffusion length ($h$) in the a-b plane. If the delamination sample could be used as a connector of two CC, as shown in Figure 5 (b). The oxygen diffusion would be firstly in the silver layer and secondly along the c-axial direction of YBCO layer. In silver, $h$ was about 2 mm, and the characteristic time ($\tau$) could be estimated to be about 1 h through the one dimensional diffusion model ($\tau=h^2/\pi^2 D$) [28]. In YBCO, $D$ along the c-axial direction could be estimated as $10^{-14}$ cm$^2$/s at 600°C [29], $d$ was about 2 μm and $\tau$ could be estimated to be about 100 h, which was comparable to the laser drilling method [27]. Moreover, the porous structure of CSD-YBCO thin film could own a much larger diffusion coefficient [28], which might result a more rapid oxygenation process for superconducting joint.

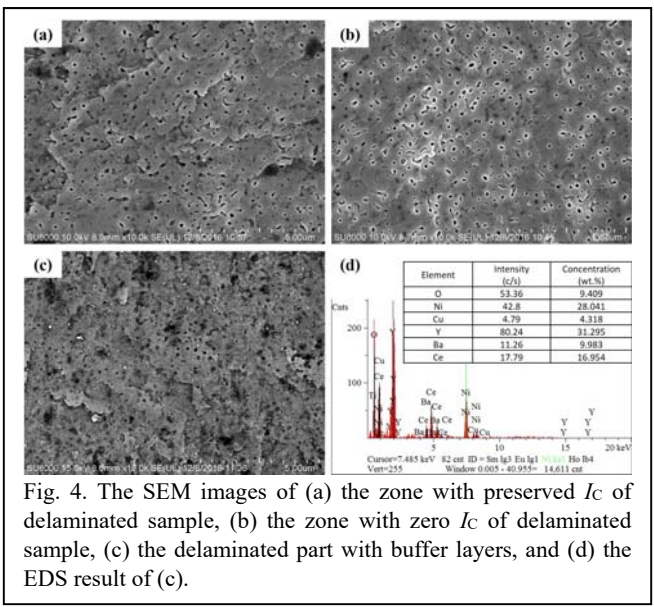

Fig. 4. The SEM images of (a) the zone with preserved $I_C$ of delaminated sample, (b) the zone with zero $I_C$ of delaminated sample, (c) the delaminated part with buffer layers, and (d) the EDS result of (c).

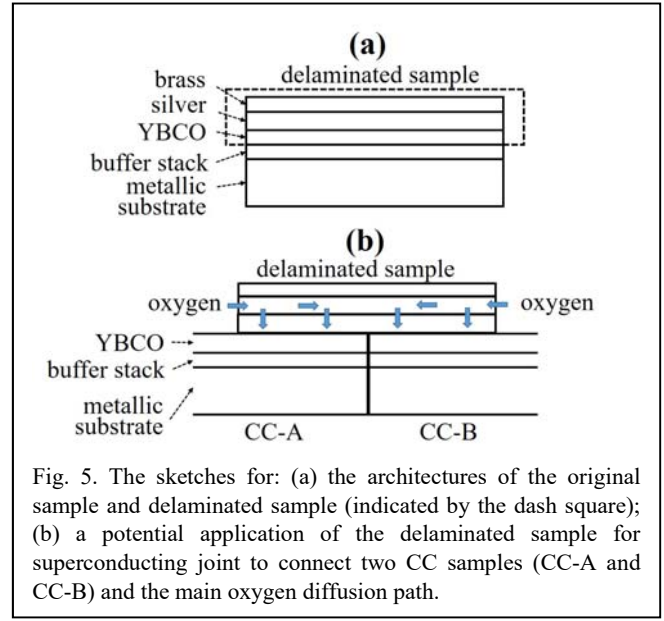

Fig. 5. The sketches for: (a) the architectures of the original sample and delaminated sample (indicated by the dash square); (b) a potential application of the delaminated sample for superconducting joint to connect two CC samples (CC-A and CC-B) and the main oxygen diffusion path.

Therefore, the CC delamination might be a beneficial phenomenon with new application potential. There were more tasks to investigate in our future study: First, the reason for $I_C$ decreasing is still not clear; Second, the post treatment such as annealing could be used to improve the property of the delaminated sample; Third, the delamination of other types of HTS CC might be different from the samples in this study.

**Conclusion**

In this study, a 60 mm long CC sample was delaminated by liquid nitrogen immersing. The $I_C$ distributions of the original sample and the delaminated sample were measured by a Hall probe scanning system. It was found that $I_C$ survived in about 2/3 of the delaminated sample, with a preserved percentage of about 45% relative to the original sample. SEM observations and EDS detections were conducted to obtain more information of the delaminated surfaces. The potential application of delaminated sample for superconducting joint technology was discussed based on the estimation of oxygen diffusion, and a much shorter oxygenation duration could be expected. More study of the $I_C$ survival in YBCO layer of the delaminated CC would be carried out in our future research.

**Acknowledgement**

This study was supported by National Natural Science Foundation of China (51402165), Fundamental Research Program of Shenzhen (JCYJ20140827160129 762), and Tribology Science Fund of State Key Laboratory of Tribology, China. The authors thank AMSC Company for the CC tape. It will be appreciated if any other groups could provide other types of HTS CC and carry out a collaborated research for further investigation on CC delamination phenomenon or superconducting joint.

**Reference**

1. D. Larbalestier, A. Gurevich, D. M. Feldmann, and A. Polyanskii, Nature 414 (6861), 368 (2001).
2. A. P. Malozemoff, D. T. Verebelyi, S. Fleshler, D. Aized, and D. Yu, Physica C 386, 424 (2003).
3. Y. Zhang, D. W. Hazelton, A. R. Knoll, J. M. Duval, P. Brownsey, S. Repnoy, S. Soloveichik, A. Sundaram, R. B. McClure, G. Majkic, and V. Selvamanickam, Physica C 473, 41 (2012).
4. D. C. van der Laan, J. W. Ekin, C. C. Clickner, and T. C. Stauffer, Supercond Sci Tech 20 (8), 765 (2007).
5. Y. Yanagisawa, H. Nakagome, T. Takematsu, T. Takao, N. Sato, M. Takahashi, and H. Maeda, Physica C 471 (15-16), 480 (2011).
6. J. E. C. Williams and E. S. Bobrov, Rev Sci Instrum 52 (5), 657 (1981).
7. T. Takematsu, R. Hu, T. Takao, Y. Yanagisawa, H. Nakagome, D. Uglietti, T. Kiyoshi, M. Takahashi, and H. Maeda, Physica C 470 (17-18), 674 (2010).
8. M. Urata and H. Maeda, IEEE T Magnetics 23 (2), 1596 (1987).
9. M. Urata and H. Maeda, IEEE T Magnetics 25 (2), 1528 (1989).
10. C. Barth, N. Bagrets, K. P. Weiss, C. M. Bayer, and T. Bast, Supercond Sci Tech 26 (5) (2013).
11. V. Arp, J Appl Phys 48 (5), 2026 (1977).
12. M. Oomen, W. Herkert, D. Bayer, P. Kummeth, W. Nick, and T. Arndt, Physica C 482, 111 (2012).
13. N. Cheggour, J. W. Ekin, C. C. Clickner, D. T. Verebelyi, C. L. H. Thieme, R. Feenstra, and A. Goyal, Appl Phys Lett 83 (20), 4223 (2003).
14. D. C. van der Laan and J. W. Ekin, Appl Phys Lett 90 (5) (2007).
15. I. Kesgin, N. Khatri, Y. Liu, L. Delgado, E. Galstyan, and V. Selvamanickam, Supercond Sci Tech 29 (1) (2016).
16. H. Kozuka, T. Fukui, M. Takahashi, H. Uchiyama, and S. Tsuboi, ACS Appl Mater Inter 4 (12), 6415 (2012).
17. H. Kozuka, T. Fukui, and H. Uchiyama, J Sol-Gel Sci Tech 67 (2), 414 (2013).
18. S. R. Foltyn, L. Civale, J. L. Macmanus-Driscoll, Q. X. Jia, B. Maiorov, H. Wang, and M. Maley, Nature Materials **6** (9), 631 (2007).
19. S. Otten, A. Kario, A. Kling, and W. Goldacker, Supercond Sci Tech 29 (12), 125003 (9 pp.) (2016).
20. K. Higashikawa, M. Inoue, T. Kawaguchi, K. Shiohara, K. Imamura, T. Kiss, Y. Iijima, K. Kakimoto, T. Saitoh, and T. Izumi, Physica C 471 (21-22), 1036 (2011).
21. K. Higashikawa, K. Shiohara, M. Inoue, T. Kiss, Takato Machi, N. Chikumoto, S. Lee, K. Tanabe, T. Izumi, and H. Okamoto, IEEE T Appl Supercond 22 (3) (2012).
22. C. Jooss, J. Albrecht, H. Kuhn, S. Leonhardt, and H. Kronmuller, Rep Prog Phys 65 (5), 651 (2002).
23. C. P. Bean, Phys Rev Lett 8 (6), 250 (1962).
24. M. Lao, J. Hecher, M. Sieger, P. Pahlke, M. Bauer, R. Hühne, and M. Eisterer, Supercond Sci Tech **30** (2) (2016).
25. W. Wu, F. Feng, Y. Zhao, X. Tang, Y. Xue, K. Shi, R. Huang, T. Qu, X. Wang, Z. Han, and J.-C. Grivel, Supercond Sci Tech 27 (5) (2014).
26. W. Goldacker, E. Mossang, M. Quilitz, and M. Rikel, IEEE T Appl Supercond 7 (2), 1407 (1997).
27. Y. Park, M. Lee, H. Ann, Y. H. Choi, and H. Lee, NPG Asia Materials 6 (2014).
28. T. Qu, Y. Xue, F. Feng, R. Huang, W. Wu, K. Shi, and Z. Han, Physica C 494, 148 (2013).
29. H. S. Kim, N. Y. Kwon, K. S. Chang, T. K. Ko, H. M. Kim, W. S. Kim, C. Park, and H. Lee, IEEE T Appl Supercond 19 (3), 2835 (2009).